\documentclass[aps, pra, preprint, amsmath, amssymb, superscriptaddress, showpacs]{revtex4-1}
\usepackage{inputenc}
\usepackage{mwe}
\usepackage{array}
\usepackage{graphicx}
\usepackage{dcolumn}
\usepackage{bm}
\usepackage{mathrsfs,amsmath} 
\usepackage{mathtools}
\usepackage{natbib}

\usepackage{enumerate}
\usepackage[]{subfigure}

\begin{document}


\title{A molecular-dynamics approach for studying the non-equilibrium behavior of x-ray-heated solid-density matter}

\author{Malik Muhammad Abdullah}
\affiliation{ Center for Free-Electron Laser Science, DESY, Notkestrasse 85, 22607 Hamburg, Germany }
\affiliation{ The Hamburg Centre for Ultrafast Imaging, Luruper Chaussee 149, 22761 Hamburg, Germany }
\affiliation{ Department of Physics, University of Hamburg, Jungiusstrasse 9, 20355 Hamburg, Germany }

\author{Anurag}
\affiliation{ Department of Physics, Indian Insitute of Technology, Kharagpur, West Bengal, India }
\affiliation{ Center for Free-Electron Laser Science, DESY, Notkestrasse 85, 22607 Hamburg, Germany }

\author{Zoltan Jurek}%
\affiliation{ Center for Free-Electron Laser Science, DESY, Notkestrasse 85, 22607 Hamburg, Germany }
\affiliation{ The Hamburg Centre for Ultrafast Imaging, Luruper Chaussee 149, 22761 Hamburg, Germany }

\author{Sang-Kil Son}
\affiliation{ Center for Free-Electron Laser Science, DESY, Notkestrasse 85, 22607 Hamburg, Germany }
\affiliation{ The Hamburg Centre for Ultrafast Imaging, Luruper Chaussee 149, 22761 Hamburg, Germany }

\author{Robin Santra}
\affiliation{ Center for Free-Electron Laser Science, DESY, Notkestrasse 85, 22607 Hamburg, Germany }
\affiliation{ The Hamburg Centre for Ultrafast Imaging, Luruper Chaussee 149, 22761 Hamburg, Germany }
\affiliation{ Department of Physics, University of Hamburg, Jungiusstrasse 9, 20355 Hamburg, Germany }

\date{\today}            
\begin{abstract}
When matter is exposed to a high-intensity x-ray free-electron-laser pulse, the x rays excite inner-shell electrons leading to the 
ionization of the electrons through various atomic processes and creating high-energy-density plasma, i.e., warm or hot dense matter. 
The resulting system consists of atoms in various electronic configurations, thermalizing on sub-picosecond to picosecond timescales after photoexcitation. 
We present a simulation study of x-ray-heated solid-density matter. 
For this we use XMDYN, a Monte-Carlo molecular-dynamics-based code with periodic boundary conditions, 
which allows one to investigate non-equilibrium dynamics. XMDYN is capable of treating 
systems containing light and heavy atomic species with full electronic configuration space and 3D spatial inhomogeneity. 
For the validation of our approach we compare for a model system the electron temperatures and the ion charge-state distribution from XMDYN to results for the thermalized system 
based on the average-atom model implemented in XATOM, an \textit{ab-initio} x-ray atomic physics toolkit extended to include a plasma environment. 
Further, we also compare the average charge evolution of diamond with the predictions of a Boltzmann continuum 
approach. We demonstrate that XMDYN results are in good quantitative agreement with the above mentioned approaches, 
suggesting that the current implementation of XMDYN is a viable approach to simulate the dynamics of x-ray-driven non-equilibrium dynamics in solids. In order to illustrate the 
potential of XMDYN for treating complex systems we present calculations on the triiodo benzene derivative 5-amino-2,4,6-triiodoisophthalic acid (I3C), a compound of relevance of 
biomolecular imaging, consisting of heavy and light atomic species.

\end{abstract}

\maketitle

\section{\label{sec:level1}Introduction}

X-ray free-electron lasers (XFELs)~\cite{pellegrinireview2016,Christophreview2016} provide intense radiation with a pulse duration down to only tens of 
femtoseconds. The cross sections for the elementary atomic processes during x-ray--matter interactions are small. Delivering high x-ray fluence can increase 
the probabilities of photoionization processes to saturation~\cite{lindayoung2010}. Nonlinear phenomena arise because of the complex multiphoton ionization pathways 
within molecular or dense plasma environment~\cite{vinko2012,zastrau,levy,zoltan2014,tachibana2015}. Theory has a key role in revealing the importance of different 
mechanisms in the dynamics. Many models have been developed for this purpose using both particle and continuum 
approaches~\cite{hauriege2004,beata2006,peyrusse2012,Scott2001,bergh2004,zoltan2004-2,saalmann2009,caleman2011-2,haureige2012}. 
In order to give a complete description of the evolution of the atomic states in the plasma, one needs to account for the possible occurrence of all 
electronic configurations of the atoms/ions. A computationally demanding situation arises when a plasma consists of heavy atomic species~\cite{rudek2012,Fukuzawa2013}. 
For example, at a photon energy of 5.5\,keV, the number of electronic configurations accessible in a heavy atom such as xenon ($Z$=54) is about 
20 million~\cite{Fukuzawa2013}. If one wants to describe the accessible configuration space of two such atoms, one must deal 
with $(2\times10^7)^2$ = $4\times10^{14}$ electronic configurations. It is clear that following the populations of all electronic configurations in 
a polyatomic system as a function of time is a formidable task. To avoid this problem, the approximation of using superconfigurations has long been 
used~\cite{Barshalom,peyrusse2000,bauche2015}. Moreover, the approach of using a set of average configurations~\cite{chung2005,lee1987} and the approach of limiting 
the available configurations by using a pre-selected subset of configurations in predominant relaxation paths~\cite{beata2016} has been applied.

The most promising approach to address this challenge is to sample the most important 
pathways in the unrestricted polyatomic electronic configuration space. This can be realized by using a Monte-Carlo strategy, which is straightforward to implement 
in a particle approach. In the present study we simulate the effect of individual ultrafast XFEL pulses 
of different intensities incident on a model system of carbon atoms placed on a lattice and analyze the quasi-equilibrium plasma state of the material reached through 
ionization and electron plasma thermalization. In order to have a comprehensive description during electron plasma thermalization we include all possible atomic electronic 
configurations for Monte-Carlo sampling, and no pre-selection of transitions and configurations is introduced. To this end, we use 
XMDYN~\cite{xatom-xmdyn,zoltan2014,tachibana2015}, a Monte-Carlo molecular-dynamics based code.

XMDYN gives a microscopic description of a polyatomic system, and phenomena such as sequential multiphoton ionization~\cite{lindayoung2010,rudek2012}, 
nanoplasma formation~\cite{tachibana2015}, thermalization of electrons through collisions and thermal emission~\cite{tachibana2015} emerge as an outcome of a simulation.
Probabilities of transitions between atomic states are determined by cross-section and rate data that are calculated by XATOM~\cite{xatom-xmdyn,sangyoulinda,sangkil2012}, 
a toolkit for x-ray atomic physics. In XMDYN individual ionization and relaxation paths are generated via a Monte-Carlo algorithm. A recent extension of XMDYN to periodic 
boundary conditions allows us to investigate bulk systems~\cite{abdullah2015,abdullah2016}. 

To validate the XMDYN approach towards a free-electron thermal equilibrium, we use an average-atom (AA) extension of XATOM~\cite{sonPhysRevX}, which is 
based on concepts of average-atom models used in plasma physics ~\cite{Rozsnyai1972,Liberman1979,Perrot1982,Peyrusse2006,WILSON2006658}. 
AA gives a statistical description of the behavior of atoms immersed in a plasma environment. It calculates 
plasma properties such as ion charge-state populations and plasma electron densities for a system with a given temperature. We compare the electron temperatures 
and ion charge-state distributions provided by XMDYN and AA. We also make a comparison between predictions for the ionization dynamics in irradiated diamond obtained 
by the XMDYN particle approach and results from a Boltzmann continuum approach published recently~\cite{beata2016}. With these comparisons, we demonstrate the potential 
of the XMDYN code for the description of high-energy-density bulk systems in and out of equilibrium.

Finally, we consider a complex system of 5-amino-2,4,6-triiodoisophthalic acid (I3C in crystalline form) consisting of heavy and light atomic species. We show the 
evolution of average atomic charge states and free electron thermalization. We demonstrate that XMDYN can simulate the 
dynamics of x-ray-driven complex matter with all the possible electronic configurations without pre-selecting any pathways in the electronic configuration space.

\section{\label{sec:level1}THEORETICAL BACKGROUND}
\subsection{\label{sec:level2}XMDYN: Molecular dynamics with super-cell approach}
XMDYN~\cite{xatom-xmdyn} is a computational tool to simulate the dynamics of matter exposed to high-intensity x rays. 
A hybrid atomistic approach~\cite{zoltan2004-2,xatom-xmdyn} is applied where 
neutral atoms, atomic ions and ionized (free) electrons are treated as classical particles, 
with defined position and velocity vectors, charge and mass. 
The molecular-dynamics (MD) technique is applied to calculate the real-space dynamics of these particles by solving the classical equations of motion numerically. 
XMDYN treats only those orbitals as being quantized that are occupied in the ground state of the neutral atom.
It keeps track of the electronic configuration of all the atoms and atomic ions. 
XMDYN calls the XATOM toolkit on the fly, which provides rate and cross-section data of x-ray-induced 
processes such as photoionization, Auger decay, and x-ray fluorescence, for all possible electronic configurations accessible during intense x-ray exposure. 
Probabilities derived from these parameters are then used in a  Monte-Carlo algorithm to generate a realization of the stochastic inner-shell dynamics. XMDYN 
includes secondary (collisional) ionization and recombination, the two most important processes occurring due to an environment. XMDYN has been validated 
quantitatively against experimental data on finite samples calculated within open boundary conditions ~\cite{zoltan2014,tachibana2015}. 

Our focus here is the bulk properties of highly excited matter. 
XMDYN uses the concept of periodic boundary condition (PBC) to simulate bulk behavior~\cite{abdullah2015,abdullah2016}. In the PBC concept, we calculate the irradiation-induced 
dynamics of a smaller unit, called a super-cell. A hypothetical, infinitely extended system is constructed as a periodic extension of the super-cell. 
The Coulomb interaction is calculated for all the charged particles inside the super-cell within the minimum image 
convention~\cite{metropolis}. Therefore, the total Coulomb force acting on a charge is given by the interaction with other charges within its well-defined neighborhood 
containing also particles of the surrounding copies of the super-cell.

\subsection{\label{sec:supercell}Impact ionization and recombination}
While core excited states of atoms decay typically within ten or less femtoseconds, electron impact ionization and recombination events occur throughout the thermalization process 
and are in dynamical balance in thermal equilibrium. The models used in this study consider these 
processes on different footing that we overview in this section. Within the XMDYN particle approach, electron impact ionization is not a stochastic process (i.e., no random number is 
needed in the algorithm), but it depends solely on the real space dynamics (spatial location and velocity) of the particles and on the cross section.
When a classical free electron is close to an atom/ion, its trajectory is extrapolated 
back to an infinite distance in the potential of the target ion by using energy and angular momentum conservation. Impact ionization occurs only if the impact parameter at infinity is 
smaller than the radius associated with the total electron impact ionization cross section. The total cross section is a sum of partial cross sections 
evaluated for the occupied orbitals, using the asymptotic kinetic energy of the impact electron. In the case of an ionization event the orbital to be ionized is chosen randomly, according 
to probabilities proportional to the subshell partial cross sections. XMDYN uses the binary-encounter-Bethe (BEB) cross sections~\cite{kimandrudd1994} supplied with atomic 
parameters calculated with XATOM.
Similarly, in XMDYN recombination is a process that evolves through the classical dynamics of the particles. XMDYN identifies for the ion that has the strongest Coulomb potential 
for each electron and calculates for how long this condition is fulfilled. Recombination occurs when an electron remains around the same ion for $n$ full periods 
(e.g., $n=1$)~\cite{xatom-xmdyn,georgescu07}. While recombination can be identified based on this definition, the electron is still kept classical if its classical orbital energy is higher than the  
orbital energy of the highest considered orbital $i$  containing a  vacancy. When the classical binding becomes stronger, the classical electron is removed and the occupation number 
of the corresponding orbital is incremented by one. Although treating recombination the above way is somewhat phenomenological (e.g., no cross section derived from inverse processes is used), 
in particle simulations similar treatments are common~\cite{Christian2005,Edward2013,georgescu07}. This process corresponds to three-body (or many-body) recombination as energy of electrons is 
transferred to other plasma electrons leading to the recombination event. The three-body recombination  is the predominant recombination channel in a warm-dense environment. 

\subsection{\label{sec:supercell}Electron plasma analysis}
Electron plasma is formed when electrons are ejected from atoms in ionization events and stay among the ions through an extensive period as, e.g., in bulk matter.  
The plasma dynamics are governed not only by the Coulomb interaction between the particles but also by collisional ionization, recombination, and so on. 
XMDYN follows the system from the very first photoionization event through non-equilibrium states until free electron thermalization is reached asymptotically. In order to quantify the 
equilibrium properties reached, we fit the plasma electron velocity distribution using a Maxwell-Boltzmann distribution,
\begin{equation}
 f(v) = \sqrt{\left(\frac{1}{2\pi T}\right)^3} 4\pi v^2e^{-\frac{v^2}{2T}},
\end{equation}
where $T$ represents the temperature (in units of energy), and $v$ is the electron speed. 
Atomic units are used unless specified. With the function defined in Eq.~(1) we fit the temperature, which is used later to compare with equilibrium-state calculations. 
￼
￼
%
\section{\label{sec:level1}Validation of the methodology}

\begin{figure}
\includegraphics[width=7.0cm,height=15cm]{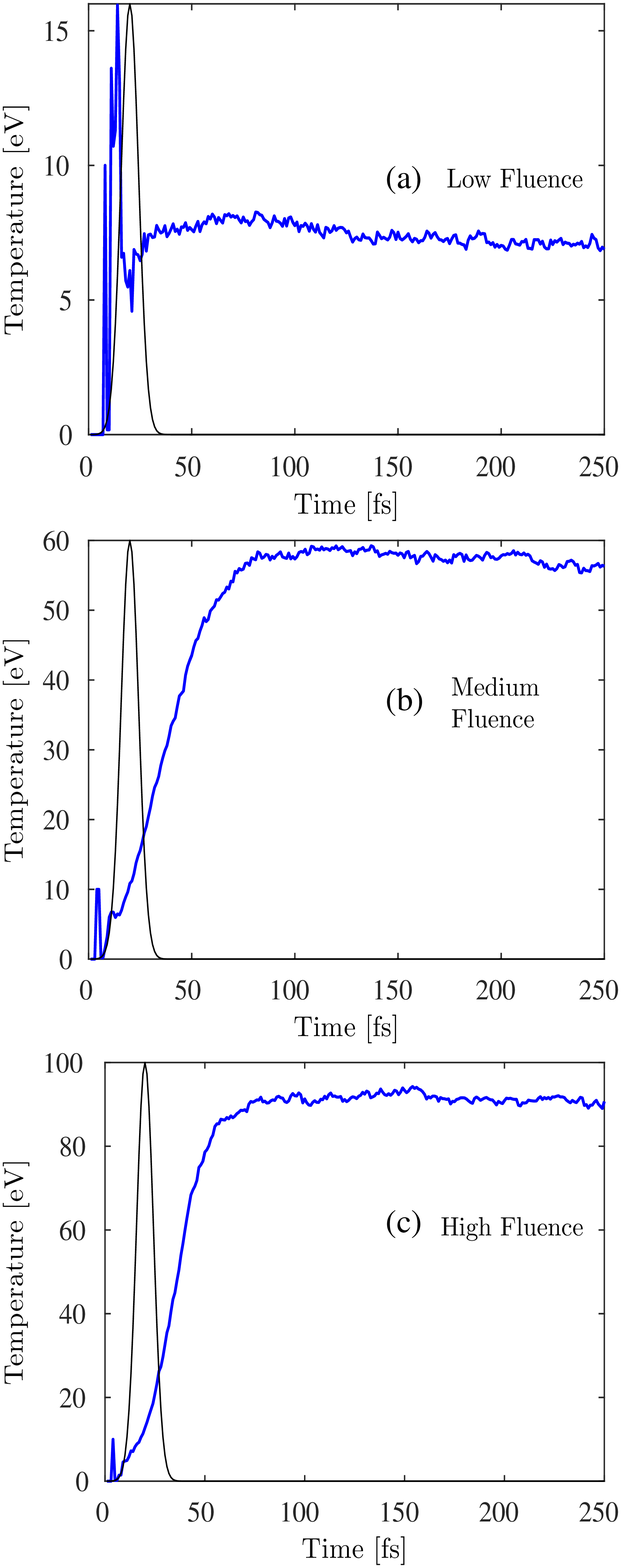}
\caption{Time evolution of the temperature of the electron plasma within XMDYN simulation during and after x-ray irradiation at different fluences: (a) 
 $\mathcal{F}_\text{low}\,=\,$6.7$\times10^{9}\,\rm{ph/}\rm{\mu m^{2}}$, (b) $\mathcal{F}_\text{med}\,=\,$1.9$\times10^{11}\,\rm{ph/}\rm{\mu m^{2}}$ and
(c) $\mathcal{F}_\text{high}\,=\,$3.8$\times10^{11}\,\rm{ph/}\rm{\mu m^{2}}$. In all three cases, the pulse duration is 10 fs FWHM; the pulse was 
centered at 20 fs, and the photon energy is 1 keV. The black curve represents the Gaussian temporal envelope. Note that in all cases equilibrium is reached within 100 fs after the pulse. }
\label{fig:temp-bulk.eps}
\end{figure}

In order to validate how well XMDYN can simulate free electron thermalization dynamics, we compare AA, where full thermalization is assumed, and XMDYN 
after reaching a thermal equlibrium. We first consider a model system consisting of carbon atoms. For a reasonable comparison of the results from XMDYN and AA, 
one should choose a system that can be addressed using 
both tools. AA does not consider any motion of atomic nuclei. Therefore we had to restrict the translational motion of atoms and atomic ions in XMDYN simulations as well. 
In order to do so, we set the carbon mass artificially so large that atomic movements were negligible throughout the calculations. Further, we increased the carbon-carbon distances 
to reduce the effect of the neighboring ions on the atomic electron binding energies. 
In XMDYN simulations, we chose a super-cell of 512 carbon atoms arranged in a diamond structure, but 
with a 13.16\,\AA\ lattice constant (in case of diamond it is 3.567\,\AA). The number density of the carbon atoms is $\rho_{0}=3.5\times10^{-3} \rm{\AA{}^{-3}}$, which 
corresponds to a mass density of $0.07 \rm{g/ cm^{3}}$. Plasma was generated by choosing different irradiation conditions typical at XFELs. 
Three different fluences, $\mathcal{F}_\text{low}\,=\,$6.7$\times10^{9}\,\rm{ph/}\rm{\mu m^{2}}$ ,
 $\mathcal{F}_\text{med}\,=\,$1.9$\times10^{11}\,\rm{ph/}\rm{\mu m^{2}}$, and $\mathcal{F}_\text{high}\,=\,$3.8$\times10^{11}\,\rm{ph/}\rm{\mu m^{2}}$, were considered. 
In all three cases the photon energy and pulse duration were 1 keV and 10 fs (full width at half maximum), respectively.
From XMDYN plasma simulations shown in Fig.~\ref{fig:temp-bulk.eps}, the time evolution of the temperature of the electron plasma is analyzed by fitting to Eq.~(1). Counterintuitively,  
right after photon absorption has finished, the temperature is still low, and then it gradually increases although no more energy is pumped into the system. The reason is that 
during the few tens of femtoseconds irradiation the fast photoelectrons are not yet part of the free electron thermal distribution; initially only the low-energy secondary electrons and Auger 
electrons that have lost a significant part of their energy in collisions determine the temperature. The fast electrons thermalize on longer timescales as shown in 
Figs.~\ref{fig:temp-bulk.eps}(b) and (c), contributing to the equilibrated subset of electrons. In all cases equilibrium is reached within 100 fs after the pulse.

\begin{figure}
\includegraphics[width=7.0cm]{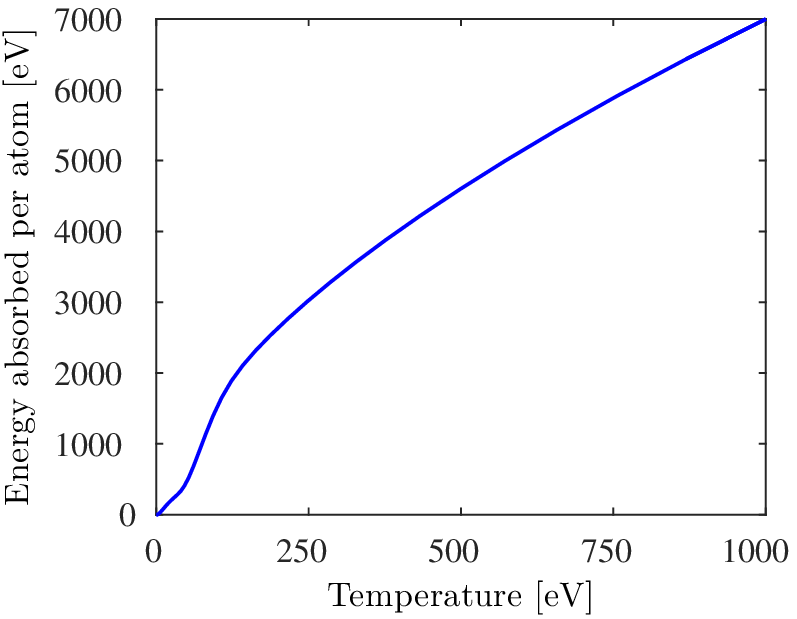}
\caption{Relation between plasma temperature and energy absorbed per atom in AA calculations for a carbon system of mass density $0.07 \rm{g/ cm^{3}}$.}
\label{fig:Energ-absorb-Bulk}
\end{figure}

AA calculates only the equilibrium properties of the system, which means that it does not consider the history of the system's evolution through non-equilibrium states. 
We first calculate the total energy per atom, $E(T)$, as a function of temperature $T$ within a carbon system of density $\rho_{0}$.
\begin{equation}
E(T) = \sum_p \varepsilon_p \tilde{n}_p(\mu,T) \int_{r \leq r_s} \!\!\! d^3 r \, \left| \psi_p(\mathbf{r}) \right|^2,
\end{equation}
where $p$ is a one-particle state index, $\varepsilon_p$ and $\psi_{p}$ are corresponding orbital energy and orbital, and $\tilde{n}_p$ stands for the fractional occupation numbers 
at chemical potential $\mu$. Details are found in Ref. [\onlinecite{sonPhysRevX}]. In this way we obtain a relation between the average energy absorbed per atom, 
$\Delta{E}=E(T)-E(0)$, and the electron temperature (see Fig.~\ref{fig:Energ-absorb-Bulk}). 
From XMDYN the average number of photoionization events per atom, ${n_{\mathrm{ph}}}$, is available for each fluence point, and therefore 
the energy absorbed on average by an atom is known (= ${n_{\mathrm{ph}}}\times\rm{\omega_{\mathrm{ph}}}$, where $\rm{\omega_{\mathrm{ph}}}$ is the photon energy).  
Using this value we can select the corresponding temperature that AA yields. This temperature is compared with that fitted from XMDYN simulation.
All these results are in reasonable agreement, as shown in Table~\ref{table:result_table}. Later we use this temperature for calculating the charge-state distributions. 

\begin{table*}
\centering
\begin{tabular}{|p{5.5cm}|p{3.5cm}|p{3.5cm}|p{3.5cm}|}
\hline
\textbf{Parameters}& \textbf{Low fluence} & \textbf{Medium fluence}& \textbf{High fluence}\\
\hline
\hline
Fluence (ph/$\rm{\mu m^2}$) & $6.7\times10^9$ & $1.9\times10^{11}$ & $3.8\times10^{11}$\\
Energy absorbed per atom (eV) & $29$ & $665$ & $1170$\\
XMDYN temperature (eV) & $7$ & $57$ & $91$\\
AA temperature (eV) & $6$ & $60$ & $83$\\
\hline
\end{tabular}
\caption{Final temperatures obtained from XMDYN runs after 250 fs propagation and from AA calculations. XMDYN temperatures are obtained from fitting using Eq.~(1), 
while AA temperatures are obtained from the absorbed energy--temperature relation (Fig.~\ref{fig:Energ-absorb-Bulk}).}
\label{table:result_table} 
\end{table*}

\begin{figure}
\includegraphics[width=0.5\textwidth]{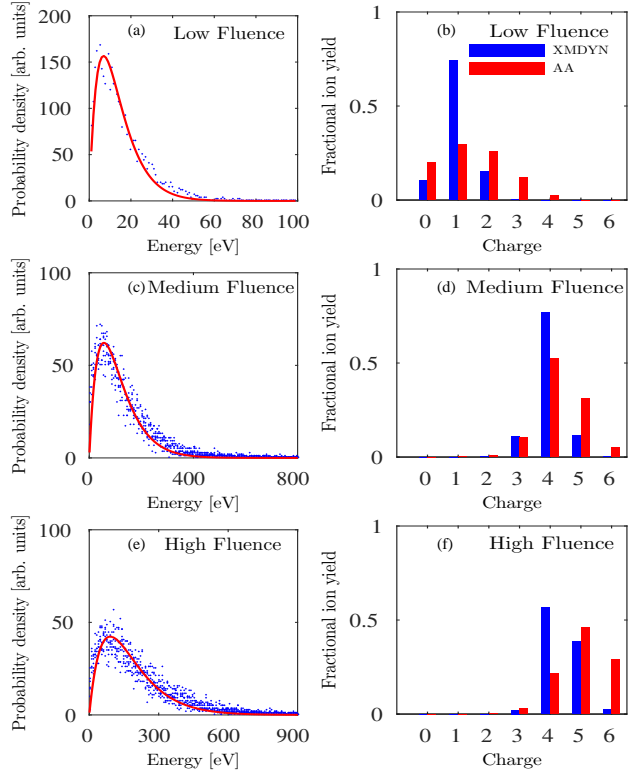}
\caption{ Kinetic-energy distribution of the electron plasma and charge-state distributions from AA and XMDYN simulations (250 fs after the irradiation) for 
the low fluence (a,b), the medium fluence (c,d), and the high fluence (e,f).}
\label{fig:frac-yield-elec-dis}
\end{figure}

Figure~\ref{fig:frac-yield-elec-dis} shows the kinetic-energy distribution  of the electron plasma (in the left panels) and the 
charge-state distributions (in the right panels) for the three different fluences. The charge-state distributions obtained from XMDYN at the final timestep (250 fs) are compared 
to those obtained from AA at the temperatures specified in Table~\ref{table:result_table}. Although similar charge states are populated using the two approaches, 
differences can be observed: AA yields consistently higher ionic charges than XMDYN (20\%--30\% higher average charges) for the cases investigated. 

This is probably for the following reasons. XMDYN calls XATOM on the fly to calculate re-optimized orbitals for 
each electronic configuration. In this way XMDYN accounts for the fact that ionizing an ion of charge $Q$ costs less energy than ionizing an ion of charge $Q+$1. However, 
in the current implementation of AA, this effect is not considered. At a given temperature, AA uses the same orbitals (and, therefore, the same orbital energies) irrespective of 
the charge state. A likely consequence is that AA gives more population to higher charge states, simply because their binding 
energies are underestimated. That could also be the reason why AA produces wider charge-state distributions and predicts a somewhat higher average charge than XMDYN does. 
The other reason for the discrepancies could be the fact that XMDYN treats only those orbitals as being quantized that are occupied in the ground state of the neutral atom. For  
carbon, these are the $1s, 2s$, and $2p$ orbitals. All states above are treated classically in XMDYN, resulting in a continuum of bound states. As a consequence, the density 
of states is different and it may yield different orbital populations and therefore different charge-state distributions. Moreover, while free-electron thermalization has been ensured 
the bound electrons are not necessarily fully thermalized in XMDYN. In spite of the discrepancies observed, XMDYN and AA 
equilibrium properties are in reasonably good agreement.

\begin{figure*}
\includegraphics[width=7.25cm,height=5.75cm]{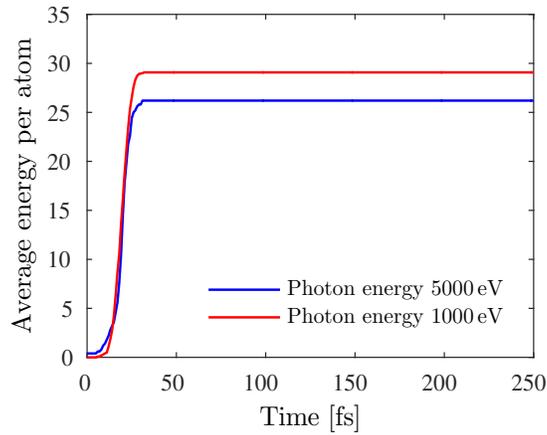}
\caption{Average energy absorbed per atom within diamond irradiated with a Gaussian pulse of hard and soft x rays 
of $\omega_{\rm{ph}}$ = 5000 eV,  $I_\text{max} = 10^{18}\, \rm{W/cm^2}$ and  $\omega_{\rm{ph}}$ = 1000 eV, $I_\text{max} = 10^{16}\, \rm{W/cm^2}$, respectively. In both cases, 
a pulse duration of 10 fs FWHM was used.}
\label{fig:energ-absorb}
\end{figure*}

\begin{figure*}
\includegraphics[width=14.0cm,height=5.6293cm]{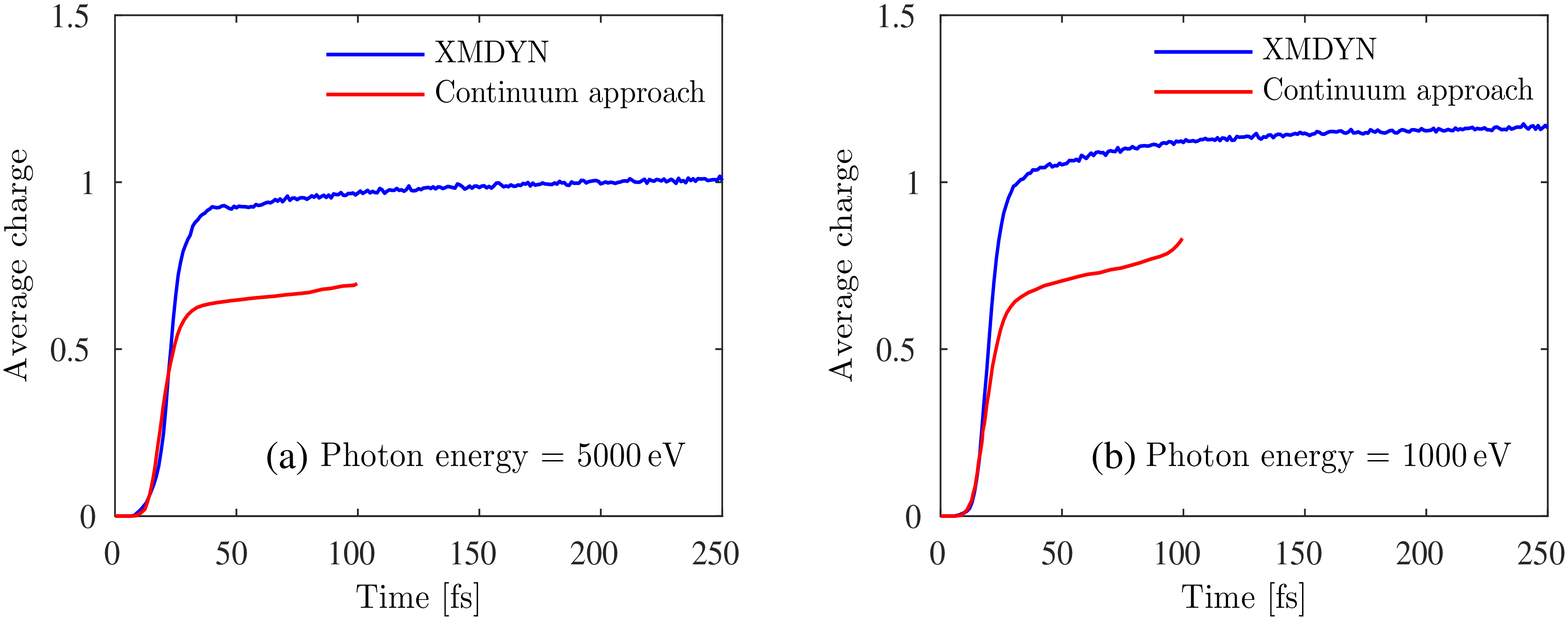}
\caption{Average charge within diamond irradiated with a Gaussian pulse of hard and soft x rays 
of (a) $\omega_{\rm{ph}}$ = 5000 eV,  $I_\text{max} = 10^{18}\, \rm{W/cm^2}$ and (b) $\omega_{\rm{ph}}$ = 1000 eV, $I_\text{max} = 10^{16}\, \rm{W/cm^2}$, respectively. In both cases, 
a pulse duration of 10 fs FWHM was used.}
\label{fig:avg-ch}
\end{figure*}

We also performed simulations under the conditions that had been used in a recent publication using a continuum approach~\cite{beata2016}. In these simulations, we do not restrict nuclear motions. 
A Gaussian x-ray pulse of 10 fs FWHM was used. The intensities considered lie within the regime typically used for high-energy-density experiments : 
$I_\text{max} = 10^{16}\, \rm{W/cm^2}$ \, for \, $\omega_{\rm{ph}}$ = 1000\,eV, and $I_\text{max} = 10^{18}\, \rm{W/cm^2}$ for  $\omega_{\rm{ph}}$ = 5000\,eV. 
We employed a super-cell of diamond (mass density = $3.51\,\rm{g/cm^{3}}$) containing 1000 carbon atoms within the PBC framework. In this study, 25 different Monte-Carlo realizations 
were calculated and averaged for each irradiation case in order to improve the statistics of the results. For a system of 1000 carbon atoms each XMDYN trajectory takes 45 minutes of 
runtime. The average energy absorbed per atom [Fig.~\ref{fig:energ-absorb}] is  $\sim{28}$\,eV and $\sim{26}$\,eV, respectively, for the 1000-eV and 5000-eV photon-energy cases, in 
agreement with Ref. [\onlinecite{beata2016}]. Figure~\ref{fig:avg-ch} shows the time evolution of the average charge for the two different photon energies. 
Average atomic charge states of +1.1 and +0.9, respectively, were obtained long after the pulse was over. 
Although the rapid increase of the average ion charge is happening on very similar times, the charge values at the end of the calculation are 
30\% and 40\% higher than those in Ref. [\onlinecite{beata2016}] for the 1000-eV and 5000-eV cases, respectively [Fig.~\ref{fig:avg-ch}(a,b)].

\begin{figure}
\includegraphics[width=7.25cm,height=5.75cm]{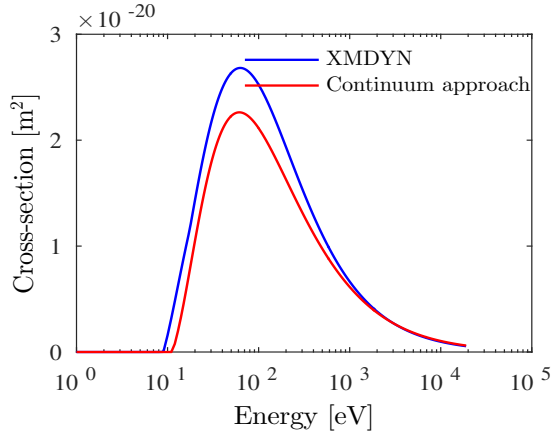}
\caption{Comparison of impact ionization cross sections for neutral ground-state carbon used in the current work within XMDYN based on the BEB formula~\cite{kimandrudd1994}, 
and the cross sections used in the continuum approach of Ref. [\onlinecite{beata2016}] based on experimental data.}
\label{fig:xmdyn-continum-cs}
\end{figure}

We can name two reasons that can cause such differences in the final charge states. One is that two different formulas for the total impact ionization cross section were used in the two approaches.
In Ref. [\onlinecite{beata2016}] the cross sections are approximated from experimental ground state atomic and ionic data~\cite{lennonetal1988}, while XMDYN employs the semi-empirical BEB formula taking 
into account state-specific properties. 
Figure~\ref{fig:xmdyn-continum-cs} compares these cross sections for neutral carbon atom. It can be seen that the cross section and, therefore, the rate of the ionization used by XMDYN are larger, 
which can shift the final average charge state higher as well. The second reason is the evaluation of the three-body recombination cross section. In Ref. [\onlinecite{beata2016}] recombination is defined using 
the principle of microscopic reversibility which states that the cross section of impact ionization can be used to calculate the recombination rate~\cite{vikrant2016}. 
In the current implementation of the Boltzmann code the two-body distribution 
function is approximated using one-body distribution functions in the evaluation of the rate for three-body recombination, whereas in XMDYN correlations at all levels are naturally captured 
within the classical framework due to the explicit calculation of the microscopic electronic fields. 

\section{\label{sec:level1}Application}

In order to demonstrate the capabilities of XMDYN we investigate the complex system of crystalline form I3C (chemical composition: $\rm{C_8H_4I_3NO_4\cdot H_2O}$)~\cite{Beck:ba5137} irradiated by 
intense x rays. I3C contains the heavy atomic species iodine, which makes it a good prototype for investigations of experimental phasing methods based on anomalous 
scattering~\cite{Guss806,Hendrickson51,sangkilprl,sangkil2013,Galli2015,Galli2015-2}. We considered pulse parameters used at an imaging experiment recently performed at the Linac Coherent 
Light Source (LCLS) free-electron laser ~\cite{I3C-team}. The photon energy was 9.7\,keV and the pulse duration was 10\,fs FWHM. Two different fluences were considered in the 
simulations, $\mathcal{F}_{\rm{high}}\,=\,$1.0$\times10^{13}\,\rm{ph/}\rm{\mu m^{2}}$ (estimated to be in the center of the focus) and its half value  
$\mathcal{F}_{\rm{med}}\,=\,$5.0$\times10^{12}\,\rm{ph/}\rm{\mu m^{2}}$. In these simulations, we do not restrict nuclear motions.

The computational cell used in the simulations contained 8 molecules of I3C (184 atoms in total). The time propagation ends 250\,fs after the pulse. For the analysis 50 XMDYN trajectories are calculated 
for both fluence cases. These trajectories sample the stochastic dynamics of the system without any restriction of the electronic configuration space that possesses $(2.0\times10^{7})^{24}$ 
possible configurations considering the subsystem of the 24 iodine atoms only. The calculation of such an XMDYN trajectory takes approximately 150 minutes on a Tesla M2090 GPU while the same 
calculation takes 48 hours on Intel Xenon X5660 2.80GHz CPU (single core).

\begin{figure*}
\includegraphics[width=14.5cm,height=5.75cm]{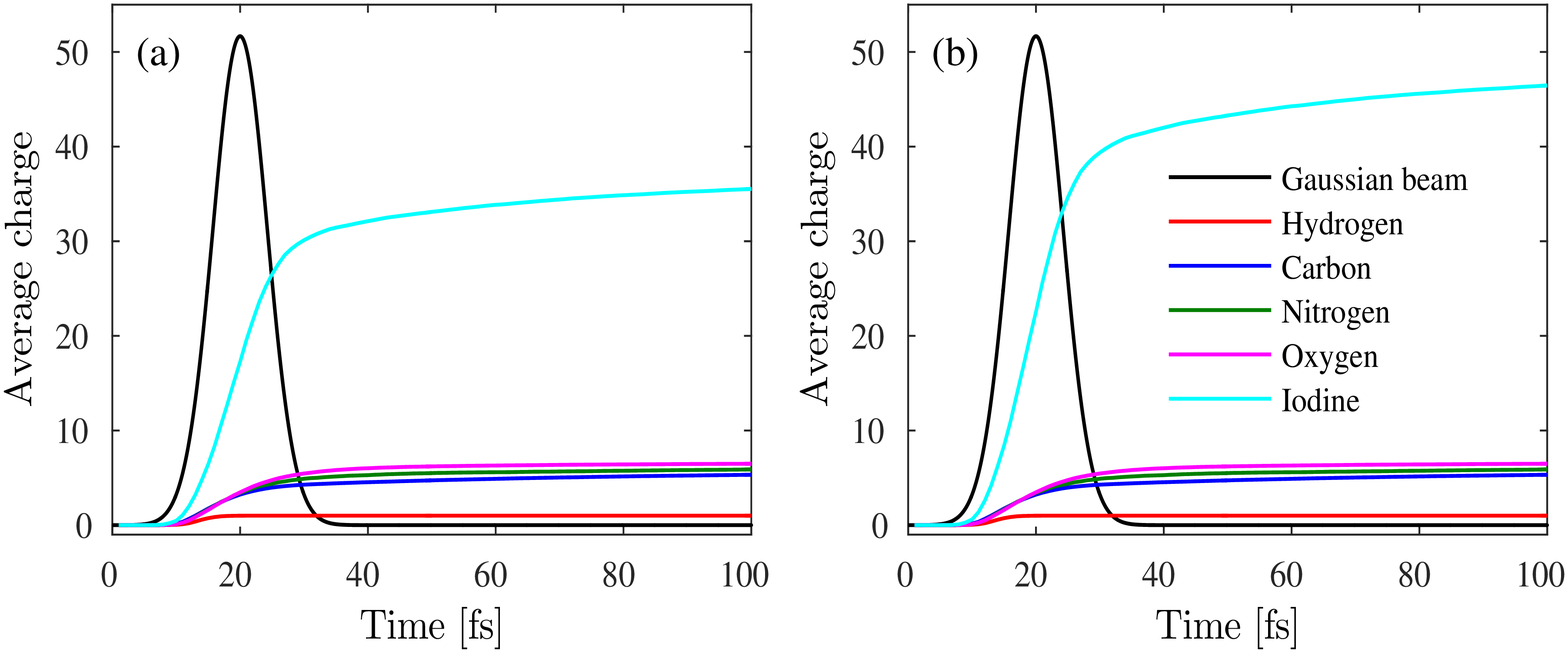}
\caption{Average atomic charge in I3C as a function of time for (a) $\mathcal{F}_{\rm{med}}\,=\,$5.0$\times10^{12}\,\rm{ph/}\rm{\mu m^{2}}$ and (b) 
$\mathcal{F}_{\rm{high}}\,=\,$1.0$\times10^{13}\,\rm{ph/}\rm{\mu m^{2}}$, respectively. In both cases, a pulse duration of 10\,fs FWHM was used. The black curve represents the 
Gaussian temporal envelope. The photon energy was 9.7\,keV.}
\label{fig:i3c_charge}
\end{figure*}

\begin{figure*}
\includegraphics[width=14.5cm,height=5.75cm]{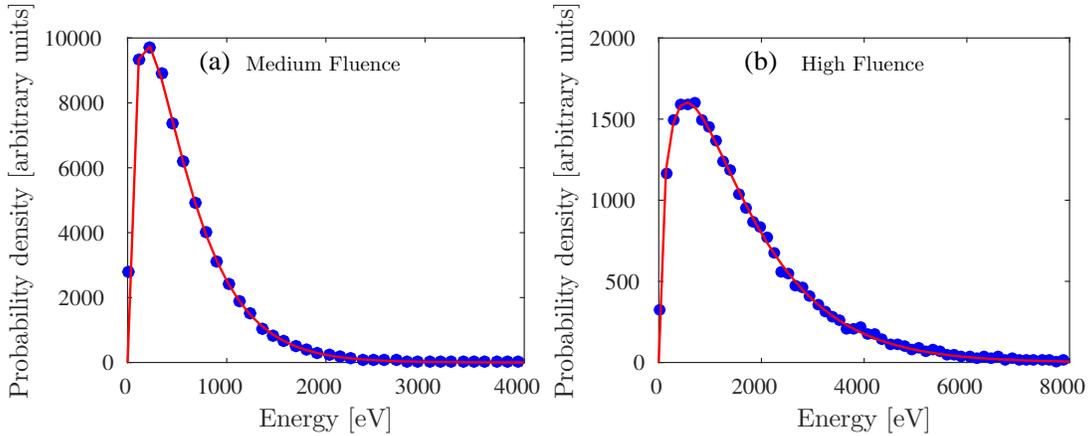}
\caption{Kinetic-energy distribution of the electron plasma in I3C from XMDYN simulations (250~fs after the irradiation) for 
(a) the medium fluence and (b) the high fluence.}
\label{fig:i3c_thermalization}
\end{figure*}

Figure~\ref{fig:i3c_charge} shows the average charge for the different atomic species in I3C as a function of time. Both fluences pump enormous energy in the system predominantly 
through the photoionization of the iodine atoms due to their large photoionization cross section. In both cases almost all the atomic electrons are removed 
from the light atoms, but mainly via secondary ionization. The ionization of iodine is very efficient: already when applying the weaker fluence $\mathcal{F}_{\rm{med}}$, the iodine atoms 
lose on average roughly half of their electrons, whereas for the high fluence case the average atomic charge goes even above +40. 
Further, we also investigate the free electron thermalization. The plasma electrons reach thermalization via non-equilibrium evolution within approximately 200\,fs. The Maxwellian 
distribution of the kinetic energy of these electrons corresponds to very high temperatures: 365\,eV for $\mathcal{F}_{\rm{med}}$ and 1\,keV for $\mathcal{F}_{\rm{high}}$ 
(see Fig.~\ref{fig:i3c_thermalization}). Hence, we have shown that XMDYN is a tool that can treat systems with 3D spatial inhomogeneity, whereas the continuum models usually deal 
with uniform or spherically symmetric samples. If the sample includes heavy atomic species, pre-selecting electronic configurations can affect the dynamics of the system. 
XMDYN allows for a flexible treatment of the atomic composition of the sample and, particularly, easy access to the electronic structure of heavy atoms with large electronic configuration space.

 
\section{\label{sec:level1}Conclusions}
We have investigated the electron plasma thermalization dynamics of x-ray-heated carbon systems using the simulation tool XMDYN and compared its predictions to two other 
conceptually different simulation methods, the average-atom model (AA) and the Boltzmann continuum approach. Both XMDYN and AA are naturally capable to address ions with 
arbitrary electronic configurations, a very common situation in high-energy-density matter generated by, e.g., high-intensity x-ray irradiation. We found very similar quasi-equilibrium 
temperatures for the two methods. 
Qualitative agreement can be observed between the predicted ion charge-state distributions, although AA tends to yield somewhat higher charges.
The reason could be that, in the current implementation, AA uses fixed atomic binding energies irrespective of the atomic electron configuration.
We have also compared results from XMDYN and the Boltzmann continuum approach for free electron thermalization dynamics of XFEL-irradiated diamond as a validation of our approach.
Thermal equilibrium of the electron plasma is reached within similar times in the two descriptions, although the asymptotic average ion charge states are somewhat different.
The discrepancy could be attributed to the different approaches for impact ionization and recombination processes in the two models and to different parametrizations used 
in the simulation. Moreover, we have considered a complex system, crystalline I3C, containing the heavy atomic species iodine. We calculated the dynamics and evolution 
of the system from an x-ray-induced non-equilibrium state to a state where the plasma electrons are thermalized and hot dense matter is formed. The atomic electronic configurations for 
iodine are taken into account in full detail. Therefore, with XMDYN the treatment of systems including heavy atomic species (exhibiting complex inner-shell relaxation pathways) is 
comprehensive and expected to be reliable. Finally, we note that, in contrast to a Boltzmann continuum approach, it is straightforward within {XMDYN} to treat spatially 
inhomogeneous systems consisting of several or even many atomic species.


\section*{Acknowledgement}
We thank Beata Ziaja for fruitful discussions about the Boltzmann continuum approach. We also thank John Spence, Richard Kirian, Henry Chapman, and 
Dominik Oberthuer, for stimulating the I3C calculations presented in this work. 
This work has been supported by the excellence cluster ``The Hamburg Center for Ultrafast Imaging (CUI): Structure, Dynamics and Control 
of Matter at the Atomic Scale'' of the Deutsche Forschungsgemeinschaft.

\bibliography{warm-dens-matter}

\end{document}